\documentclass[11pt,preprint]{aastex}

\usepackage{graphics,graphicx}
\usepackage{natbib}
\newcommand{\gtorder}{\mathrel{\raise.3ex\hbox{$>$}\mkern-14mu
            \lower0.6ex\hbox{$\sim$}}}
\newcommand{\ltorder}{\mathrel{\raise.3ex\hbox{$<$}\mkern-14mu
            \lower0.6ex\hbox{$\sim$}}}

\shorttitle{Disk Winds in Tidal Disruption Events}
\shortauthors{Miller}

\begin{document}

\title{DISK WINDS AS AN EXPLANATION FOR SLOWLY EVOLVING TEMPERATURES IN TIDAL DISRUPTION EVENTS}

\author {M. Coleman Miller\altaffilmark{1}}

\affil{
{$^1$}{Department of Astronomy and Joint Space-Science Institute, University of Maryland, College Park, MD 20742-2421 USA; miller@astro.umd.edu}\\
}

\begin{abstract}

Among the many intriguing aspects of optically discovered tidal disruption events (TDEs) is that their temperatures are lower than predicted and that the temperature does not evolve as rapidly with decreasing fallback rate as would be expected in standard disk theory.  We show that this can be explained qualitatively using an idea proposed by Laor \& Davis in the context of normal active galactic nuclei: that larger accretion rates imply stronger winds and thus that the accretion rate through the inner disk only depends weakly on the inflow rate at the outer edge of the disk.  We also show that reasonable quantitative agreement with data requires that, as has been suggested in recent papers, the characteristic radius of the tidal stream is approximately equal to the semimajor axis of the most bound orbit of the debris rather than twice the pericenter distance, which would be expected from circularization without rapid angular momentum redistribution.  If this explanation is correct, it suggests that the evolution of TDEs may test both non-standard disk theory and the details of the interactions of the tidal stream.

\end{abstract}


\section{INTRODUCTION}
\label{sec:introduction}

Tidal disruption events (TDEs), in which a star is torn apart by a massive black hole, have emerged in the last few years as excellent examples of what can be learned in time domain astronomy.  The possible scientific returns include elucidation of the powering of some active galactic nuclei (AGNs; \citealt{1975Natur.254..295H} and many subsequent papers), probes of otherwise nearly dormant massive black holes \citep{1976MNRAS.176..633F,1979PAZh....5...28L}, tests of hydrodynamics and magnetohydrodynamics of streams in strong gravity \citep{1982Natur.296..211C,1982ApJ...262..120L,1988Natur.333..523R,1989ApJ...346L..13E,1989IAUS..136..543P,1999ApJ...519..647K,2000ApJ...545..772A,2009MNRAS.392..332L,2012PhRvD..86f4026K,2013ApJ...767...25G}, and exploration of general relativistic effects such as pericenter precession and Lense--Thirring precession \citep{1994ApJ...422..508K,2012PhRvL.108f1302S,2012ApJ...749..117H,2013ApJ...775L...9D,2013MNRAS.434..909H,2014PhRvD..90f4020C,2014ApJ...783...23G,2014ApJ...784...87S,2015arXiv150104365S,2015arXiv150104635B,2015arXiv150105306G,2015arXiv150105207H}.  

A still small, but growing, number of candidate TDEs have been discovered in X-ray \citep{1988ApJ...325L..25P,1995MNRAS.273L..47B,1995A&A...299L...5G,1996A&A...309L..35B,1999A&A...350L..31G,1999A&A...343..775K,1999A&A...349L..45K,2000A&A...362L..25G,2002AJ....124.1308D,2008A&A...489..543E,2009A&A...495L...9C,2010ApJ...722.1035M,2012A&A...541A.106S}, ultraviolet \citep{1995Natur.378...39R,2004ApJ...612..690S,2006ApJ...653L..25G,2008ApJ...676..944G,2009ApJ...698.1367G}, and optical \citep{2011ApJ...741...73V,2012Natur.485..217G,2014ApJ...793...38A,2014ApJ...780...44C,2014MNRAS.445.3263H,2015ApJ...798...12V} surveys.
Among the many puzzles posed by optically discovered TDEs is that their temperatures are not as high as predicted, and that the temperatures do not appear to decrease as rapidly as expected in standard disk theory.  For example, if a black hole of mass $M$ accretes at a rate ${\dot M}$, then \citet{1973A&A....24..337S} find that the surface radiative flux at a radius $R$ is 
\begin{equation}
F=\sigma T_{\rm eff}^4={3GM{\dot M}\over{8\pi R^3}}\left[1-\left(R_{\rm in}\over R\right)^{1/2}\right]\; .
\end{equation}
Here $\sigma$ is the Stefan--Boltzmann constant, $G$ is Newton's constant, and $R_{\rm in}$ is the inner radius of the disk, which in this formulation is assumed to be torque-free.  Maximizing the flux over $R$ we find that the peak disk effective temperature $T_{\rm max}$ at a fraction $f$ of the Eddington rate ${\dot M}_{\rm Edd}=4\pi GM/(\kappa\eta c)$ is given by 
\begin{equation}
\sigma T_{\rm max}^4={3\over 2}{6^6\over{7^7}}{fG^2M^2\over{\kappa\eta cR_{\rm in}^3}}\; .
\end{equation}
Here $\kappa$ is the opacity, $\eta=L/{\dot M}c^2$ is the accretion efficiency that gives a luminosity $L$ for an accretion rate ${\dot M}$, and $c$ is the speed of light.   If we set $R_{\rm in}=6GM/c^2$, which is the Boyer--Lindquist radius of the innermost stable circular orbit for a nonrotating black hole, then
\begin{equation}
T_{\rm max}=4.8\times 10^5~{\rm K}~f^{1/4}(\kappa/0.4~{\rm cm}^2~{\rm g}^{-1})^{-1/4}(\eta/0.057)^{-1/4}(M/10^6~M_\odot)^{-1/4}
\end{equation}
(for a similar expression see, e.g., \citealt{1999ApJ...514..180U}, who also gives a somewhat lower temperature and different scaling if $R_{\rm in}$ is set by the tidal radius).  Here we have scaled to the pure hydrogen Thomson opacity and the efficiency $\eta=0.057$ for a nonrotating black hole.  Thus one would expect maximum disk temperatures $T_{\rm max}\sim{\rm few}\times 10^5$~K, and one would also expect that the temperature would decrease significantly over the course of an outburst.

Neither of these expectations is met in optically discovered TDEs.  The typical best-fit Planck temperature to the spectra is ${\rm few}\times 10^4$~K rather than ${\rm few}\times 10^5$~K \citep{2011ApJ...741...73V,2012Natur.485..217G,2014ApJ...793...38A,2014ApJ...780...44C,2014MNRAS.445.3263H,2015ApJ...798...12V}.  It has been proposed that when the fallback rate is super-Eddington there is an optically thick photosphere at many times the tidal radius, and that this could lead to characteristic temperatures $\sim 10^4$~K \citep{1997ApJ...489..573L,1998A&A...333..379U,2009MNRAS.400.2070S,2014ApJ...781...82C,2014ApJ...783...23G}; see also Figure~3 of \citet{2014arXiv1410.7772S} for a comparison with data.  However, there are now multiple events in which the best-fit temperature remains roughly constant even as the fallback rate decreases by a large factor.  For example, PS1--10jh maintains a nearly constant UV-optical spectral shape even as the integrated flux from this event drops by more than an order of magnitude (see Figure~3 of \citealt{2012Natur.485..217G}).  Similarly, ASASSN--14ae maintains a best-fit Planck temperature of $\sim 2\times 10^4$~K (uncorrected for extinction) as the integrated flux drops by one and a half orders of magnitude \citep{2014MNRAS.445.3263H}.  Standard disruption models, in which the fallback rate scales as $\sim t^{-5/3}$ (although see the caveats related to the structure of the star in \citealt{1989IAUS..136..543P} and \citealt{2009MNRAS.392..332L} and the relation between flux in a band and bolometric flux in \citealt{2009MNRAS.400.2070S} and \citealt{2011MNRAS.410..359L}) or even as rapidly as $t^{-2.2}$ if the stellar core survives \citep{2013ApJ...767...25G} suggest that the fallback rate would have dropped below Eddington after such a large drop in flux for black hole masses greater than $\sim 10^6~M_\odot$.  It is possible that the size of a reprocessing layer adjusts with the fallback rate to yield a nearly constant maximum temperature (e.g., \citealt{2014ApJ...783...23G}), but it is not clear what physics would enforce this.  Thus explanations that involve super-Eddington accretion rates and expanded, nearly spherical, photospheres might not suffice for the entire observed evolution for several of these sources.

Here we propose that both the low temperature and the slow evolution of temperature can be explained in a framework proposed by \citet{2014MNRAS.438.3024L} to explain why the maximum temperature seen in AGNs is also only ${\rm few}\times 10^4$~K over a wide range of black hole masses and accretion rates.  The idea is that line-driven winds, which \citet{2014MNRAS.438.3024L} calibrate using data from stars, remove mass from the disk.  At larger accretion rates at the outer disk boundary, more mass is removed and thus the peak temperature of the disk is reduced substantially from its standard disk value.  At smaller accretion rates, less mass is removed and the temperature reduction is not as great.  Thus the maximum achievable temperature is significantly less than it is in the standard disk picture, and the dependence of the peak disk temperature on the accretion rate is also much weaker than in the usual model.

An additional quantitative test of TDE scenarios is available through the optical and near ultraviolet spectra that have been obtained for some events.  If a star on a nearly parabolic orbit is tidally disrupted at a pericenter distance $r_p$ and the debris circularizes at the specific angular momentum of the initial orbit, then the Newtonian circularization radius is $2r_p$.  In contrast, as is shown in several recent papers \citep{2015arXiv150104635B,2015arXiv150105306G,2015arXiv150105207H,2015arXiv150104365S}, if shock cooling is inefficient then the characteristic disk radius is much larger.  For example, \citet{2015arXiv150104365S} specifically suggest that the radius is roughly equal to the semimajor axis $a_{\rm min}$ of the most bound orbit.  Given that $a_{\rm min}\sim (M/m)^{1/3}r_{\rm tide}$, where $m$ is the mass of the star prior to disruption and $r_{\rm tide}\sim (M/m)^{1/3}R_*$ for a stellar radius $R_*$ \citep{1988Natur.333..523R}, the two radii can differ by a factor of tens if $M/m\sim 10^5-10^6$.  As we show, this has significant implications for the optical and ultraviolet fluxes, although precise predictions are susceptible to uncertainties in the shape and surface density distribution of the disk as well as to the unknown extinction in the host galaxies of TDEs.

In Section~\ref{sec:methods} we review the salient aspects of the \citet{2014MNRAS.438.3024L} model and show the expected evolution of the maximum disk temperature as a function of accretion rate.  In Section~\ref{sec:results} we perform by-eye fits to data from the PS1--10jh \citep{2012Natur.485..217G} and PS1--11af \citep{2014ApJ...780...44C} candidate TDEs, which show that in the context of this model a large disk radius is favored strongly.  We present our conclusions in Section~\ref{sec:conclusions}.

\section{METHODS}
\label{sec:methods}

\subsection{Physical Principles}

Our suggestion is based on the disk wind model that \citet{2014MNRAS.438.3024L} propose to explain the ${\rm few}\times 10^4$~K maximum disk temperature seen from AGN.  Their wind loss prescriptions originate from observations of O stars, from which they derive two possible relations for the wind mass loss rate per area ${\dot\Sigma}$.  The first is:
\begin{equation}
{\dot\Sigma}=2.6\times 10^{-12}~{\rm g~cm}^{-2}~{\rm s}^{-1}~(F/F_\odot)^{1.9}\; ,
\end{equation}
where $F$ is the energy flux from a part of the disk and $F_\odot=6.3\times 10^{10}$~erg~s$^{-1}$~cm$^{-2}$ is the solar radiation flux.  We refer to this as ${\dot\Sigma}(F)$.  The second relation is
\begin{equation}
{\dot\Sigma}=2.48\times 10^{-14}~{\rm g~cm}^{-2}~{\rm s}^{-1}~(F/F_\odot)^{2.32}~(g/g_{\odot})^{-1.11}\; ,
\end{equation}
where $g$ is the surface gravity at the photosphere of part of the disk and $g_\odot=2.74\times 10^4~{\rm cm~s}^{-2}$ is the solar surface gravity.  We refer to this prescription as ${\dot\Sigma}(F,g)$.  As \citet{2014MNRAS.438.3024L} show, the assumptions of hydrostatic equilibrium and opacity dominated by electron scattering imply that $g/g_\odot=2.5\times 10^{-5}~F/F_\odot$, which means that ${\dot\Sigma}(F,g)$ reduces to ${\dot\Sigma}=3.05\times 10^{-9}~{\rm g~cm}^{-2}~{\rm s}^{-1}~(F/F_\odot)^{1.21}$.  The second relation is a better fit to O star data, but it is not clear which, if either, is applicable to accretion disks around massive black holes.  It is important to note that the power-law relation between ${\dot\Sigma}$ and $F$ cannot, of course, extend to unlimited fluxes and effective temperatures.  The true relation depends on the opacity as a function of temperature.  For example, Figure~1 of \citet{2009A&A...499..279C} and Figure~38 of \citet{2013ApJS..208....4P} display aspects of the OPAL opacities \citep{1996ApJ...464..943I} and show that above $T_{\rm eff}\sim 2\times 10^5$~K the opacity actually {\it drops} due to weakening of lines caused by high ionization.  Thus at sufficiently high ionization fractions (which could be due to locally high temperatures or, as is probably the case in stellar-mass X-ray binaries, reprocessed X-ray emission from the inner disk), the wind loss rates will be suppressed.

Because the winds from O stars are driven largely by the opacity due to the overlap of atomic lines (see the discussion in \citealt{2014MNRAS.438.3024L}), there should be a significant dependence of the wind strength on the metallicity of the gas, with higher metallicity implying stronger winds if other parameters are kept constant.  This is not explicitly included in the \citet{2014MNRAS.438.3024L} model, but the theory could be adjusted if there is enough information about the metallicity of a particular disrupted star.  We note that because metal line overlap is needed, the presence or absence of hydrogen is not expected to make a significant difference other than by allowing the heavier elements to be a larger fraction of the overall mass.  Thus the winds from the disruption of helium-rich stars, as has been proposed for PS1--10jh by \citet{2012Natur.485..217G}, would be expected to be stronger than but qualitatively similar to the winds from the disruption of stars with solar composition.

Another point to consider is whether the wind will be optically thick and, if so, how it will interact with the outgoing radiation.  To assess this we first need to determine whether atomic opacities or Thomson scattering dominate.  A useful parameter in this context is the ionization parameter $\xi\equiv 4\pi F/n_e$, where $n_e$ is the electron number density (see \citealt{1977ApJ...218...20D} for an early application to quasar spectra).  This is a measure of the number of ionizing photons per electron.  \citet{1993MNRAS.262..179M,1996MNRAS.278.1111M} found that when $\xi\ltorder 100~{\rm erg~cm~s}^{-1}$ the gas is only weakly ionized, whereas when $\xi\gtorder 5000~{\rm erg~cm~s}^{-1}$ the gas is strongly ionized, which means that Thomson scattering (in this case, rather than free-free absorption) will be the most important source of opacity.  

If the radial speed $v_r(R)\approx \left(GM/R\right)^{1/2}$, then the number density of electrons in a spherical wind composed of pure hydrogen is roughly
\begin{equation}
n_e={{\dot M}/m_p\over{4\pi R^2v_r(R)}}={{\dot M}/m_p\over{4\pi R^2(GM/R)^{1/2}}}\; .
\end{equation}
The ionization parameter is $\xi=4\pi F/n_e=4\pi(L/4\pi R^2)n_e$, so if we use $L=\eta{\dot M}c^2$ we find
\begin{equation}
\xi=\eta m_pc^2 4\pi(GM/R)^{1/2}\approx 2\times 10^6~{\rm erg~cm~s}^{-1}[(10^3GM/c^2)/R]^{1/2}\; .
\end{equation}
Thus the wind is highly ionized and Thomson scattering is the dominant source of opacity.  In contrast, if we use the \citet{1973A&A....24..337S} solution for the inner region at $R\gg R_{\rm in}$, we find $\xi\sim 200~{\rm erg~cm~s}^{-1}$ at the photosphere of the disk for $M=3\times 10^6~M_\odot$, $R=10^3GM/c^2$, and ${\dot M}={\dot M}_{\rm Edd}$.  This means that line opacities can dominate radiation transport in the disk but are unlikely to be important in the wind.

The optical depth through the wind is then 
\begin{equation}
\begin{array}{rl}
\tau&=\int_R^\infty n\sigma_T dr={\dot M\sigma_T\over{2\pi m_p(GMR)^{1/2}}}\\
&\approx (0.06/\eta)({\dot M}/{\dot M}_{\rm Edd})[(10^3~GM/c^2)/R]^{1/2}\; ,\\
\end{array}
\end{equation}
independent of the black hole mass.  The numerical coefficient is roughly unity for $\eta=0.057$, which means that for our standard parameters the optical depth could be a few for mass fallback rates of $\sim 10\times$ Eddington.  However, the very high ionization parameter implies that absorption opacities will not be important in the wind.  Thus the spectrum might be mildly Comptonized but seems unlikely to be affected otherwise.  However, as the referee pointed out, this scattering could smooth out variations in the light curve that are shorter than the diffusion time through the wind, when the wind is optically thick.  This could be hours to days in the peak fallback rate part of the disruption.

\subsection{Numerical Implementation}

For simplicity we assume that the disk has Keplerian rotation, i.e., $\Omega_k^2=GM/R^3$ at a disk radius $R$.  This requires that the disk is geometrically thin, because otherwise pressure support will change the rotation law.  The thin-disk assumption will be poor close to the hole if the local accretion rate is comparable to or larger than Eddington, but at hundreds of gravitational radii the standard \citet{1973A&A....24..337S} solution suggests that even disks with accretion rates several times Eddington will be geometrically thin.

With this assumption, manipulation of equations (26)--(28) of \citet{2014MNRAS.438.3024L} yields the differential equations to be integrated:
\begin{equation}
\begin{array}{rl}
{\partial{\dot M}\over{\partial R}}&=4\pi R{\dot\Sigma}\\
{\partial W_{R\phi}\over{\partial R}}&={{\dot M}\over{4\pi R^2}}\left(GM\over R\right)^{1/2}-{2\over R}W_{R\phi}\\
\end{array}
\end{equation}
where $W_{R\phi}$ is the vertically integrated stress and as above ${\dot\Sigma}$ is a function of $F$, where
\begin{equation}
F={3\over 4}W_{R\phi}\left(GM\over{R^3}\right)^{1/2}-{\epsilon\over 2}{GM\over R}{\dot\Sigma}\; .
\end{equation}
Here $\epsilon$, a parameter in the \citet{2014MNRAS.438.3024L} model, is the ratio of the wind speed at infinity to the escape speed from the location in the disk where the wind originated.  Note that although we treat $\epsilon$ as independent of the effective temperature, it need not be: for example, as pointed out by Nick Stone (private communication), there is some evidence from massive stars that there is a jump in wind speed when the temperature is $\sim 1-2\times 10^4$~K (e.g., \citealt{1990A&A...237..409P}).  

Solution of this system of equations requires a choice for the inner boundary condition on $W_{R\phi}$.  For simplicity we choose the zero-stress zero-spin boundary condition
\begin{equation}
W_{R\phi}=0\;{\rm at}~R_{\rm in}=6GM/c^2
\end{equation}
and verify that when ${\dot M}$ is set to zero everywhere the solution agrees with the analytical expression $W_{R\phi}=({\dot M}\Omega_k/2\pi)\left[1-(R_{\rm in}/R)^{1/2}\right]$.  Note that zero inner stress is neither physically required nor consistent with all magnetohydrodynamic simulations (e.g., \citealt{2010ApJ...711..959N}); it is simply a convenient reference case, and the results we display depend only weakly on the inner boundary condition.  For numerical reasons it is easiest to integrate outward from the inner boundary rather than inward from the outer boundary.  Thus for a given desired mass accretion rate ${\dot M}_{\rm out}$ at the outer disk boundary $R_{\rm out}$ we make a guess for ${\dot M}_{\rm in}$ at the inner boundary $R_{\rm in}$ (which we set to $6GM/c^2$ for the results shown here), integrate outward to $R_{\rm out}$ and measure ${\dot M}_{\rm out}$, and converge on the solution for ${\dot M}_{\rm in}$ by bisection.  

\begin{figure}[!htb]
\begin{center}
\plotone{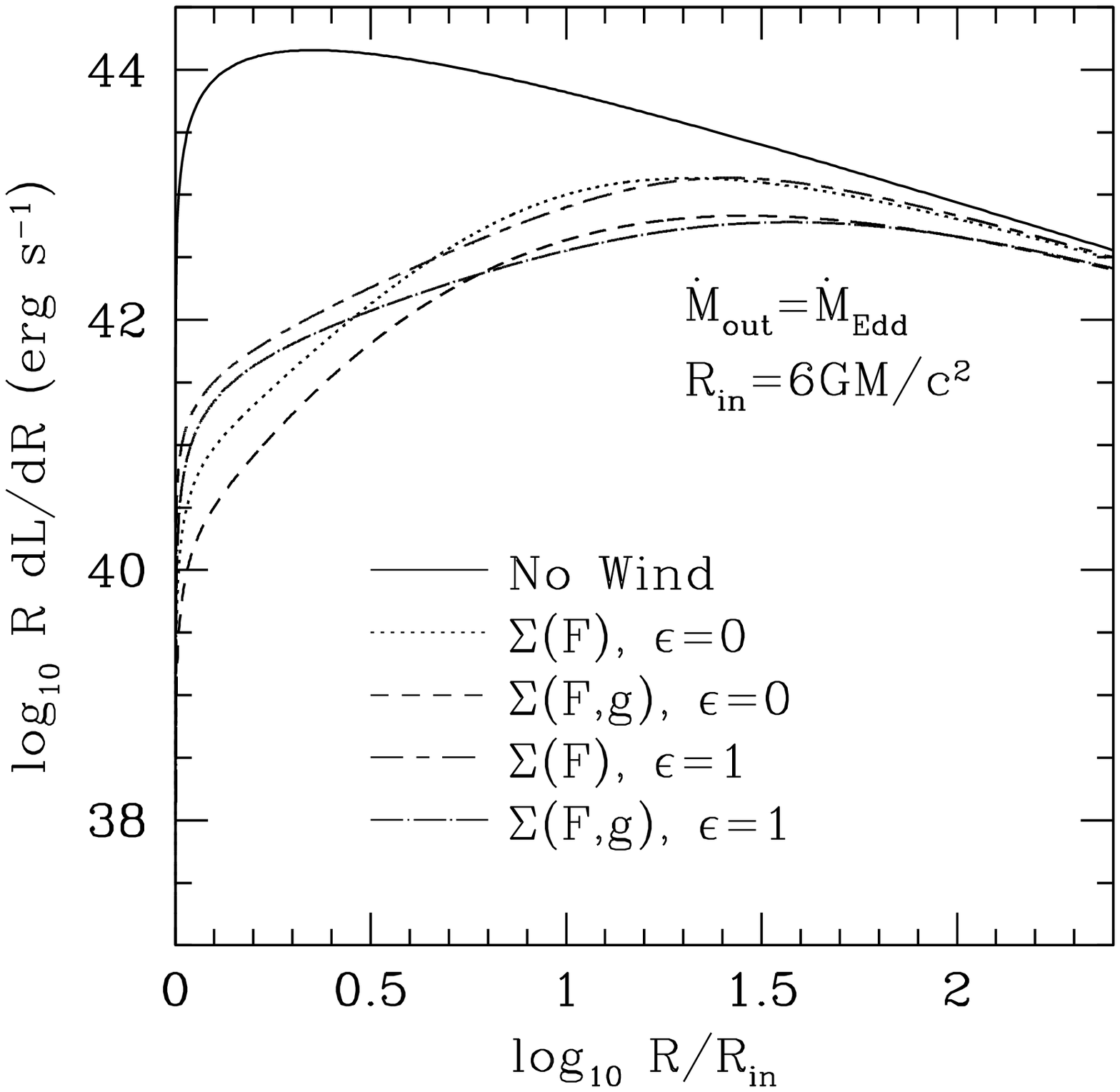}
\vskip-1.5cm
\caption{Gradient of bolometric luminosity times radius as a function of radius for models with $M=3\times 10^6~M_\odot$ and an outer accretion rate equal to the Eddington rate.  The solid line shows the standard no-wind solution, and the other line types show solutions with winds assuming  ${\dot\Sigma}={\dot\Sigma}(F)$ or ${\dot\Sigma}={\dot\Sigma}(F,g)$ and $\epsilon=0~{\rm or}~1$.  As expected, at large radii the solutions asymptote to each other, but at smaller radii the mass loss reduces the luminosity considerably in the wind models.  See Figures~4--6 of \citet{2014MNRAS.438.3024L} for similar comparisons of the accretion rate and effective temperature as a function of radius, and of the total spectrum from the disk.
}
\label{fig:dLdr}
\end{center}
\end{figure}

In Figure~\ref{fig:dLdr} we show the effect of the wind for various prescriptions ($\Sigma(F)$ or $\Sigma(F,g)$, and $\epsilon=0$ or $\epsilon=1$).  We assume that the outer accretion rate equals the Eddington rate, and track the luminosity gradient times the radius, which shows the radius at which the luminosity peaks.  We see that at large distances from the hole the solutions converge, as expected.  At smaller distances, the luminosity is less at a given radius than it would be without a wind, by an amount that depends somewhat on the prescription.  We also see that the luminosity peaks at a larger radius than it would without a wind.  One consequence of this is that the maximum temperature of the disk is lowered significantly by a wind.

\begin{figure}[!htb]
\begin{center}
\plotone{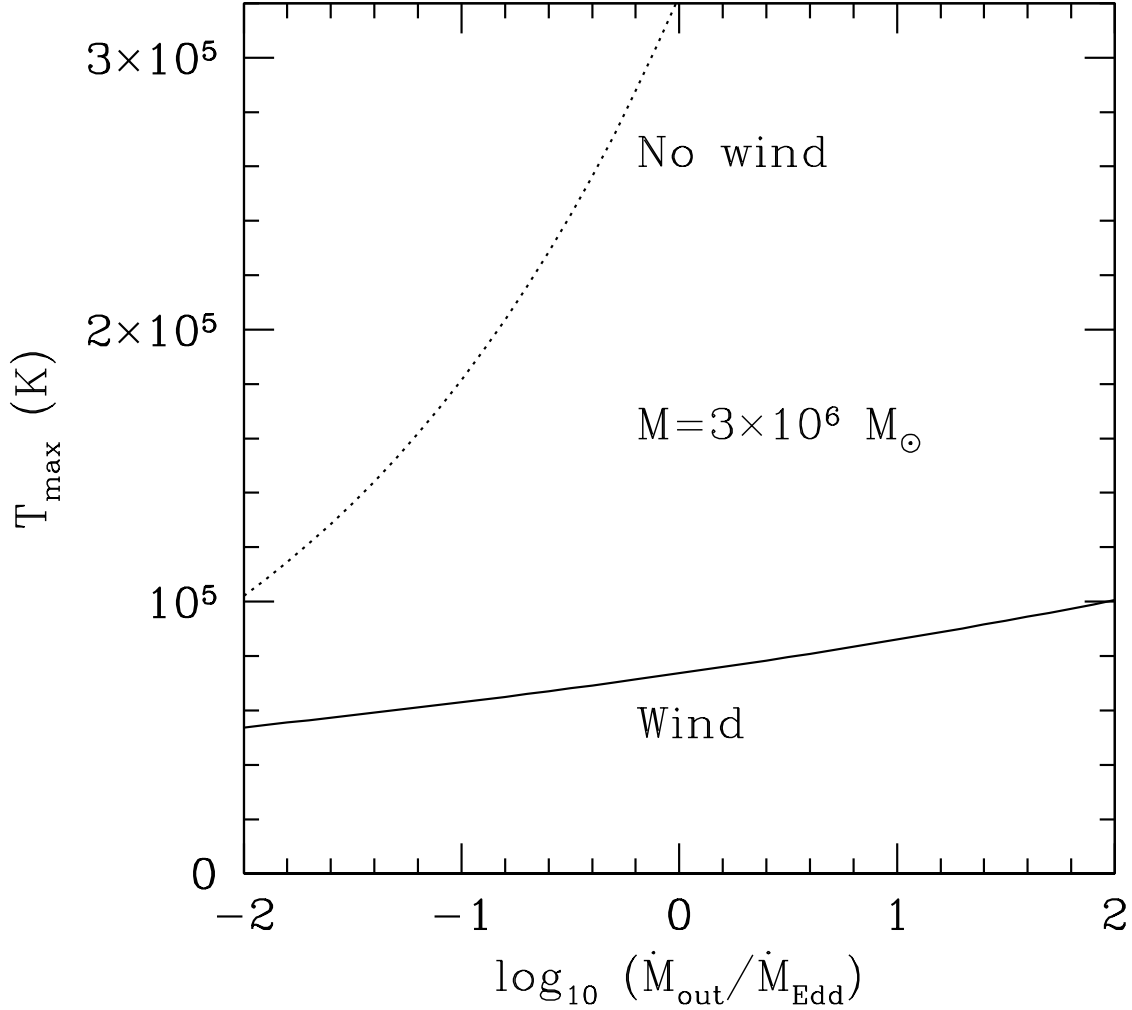}
\vskip-1.5cm
\caption{Maximum disk effective temperature vs. infall accretion rate in Eddington units at the outer edge of the disk as predicted using the \citet{2014MNRAS.438.3024L} model (solid line), with $\epsilon=0$ and ${\dot\Sigma}={\dot\Sigma}(F)$ (see Section~\ref{sec:methods} for details; other model choices give similar results).  For comparison we show the maximum temperature in the standard no-wind disk solution (dotted line).  Over a factor of $10^4$ in infall rate the maximum temperature in the wind model changes by only a factor of 1.9, compared with the factor of 10 predicted in the standard disk model.  In addition, the maximum temperature at an Eddington accretion rate for this $M=3\times 10^6~M_\odot$ black hole is only $7.4\times 10^4$~K, rather than $T_{\rm max}=3.2\times 10^5$~K as predicted for standard disks.  In both respects, the disk wind model of \citet{2014MNRAS.438.3024L} is a better fit than the no-wind model to data from optically discovered TDEs.
}
\label{fig:temp}
\end{center}
\end{figure}

This maximum temperature is plotted in Figure~\ref{fig:temp}.  We show the disk maximum temperature as a function of outer accretion rate, assuming ${\dot\Sigma}={\dot\Sigma}(F)$ and $\epsilon=0$; other choices for ${\dot\Sigma}$ and $\epsilon$ give qualitatively similar results.  The particular dependence we find is close to the $T_{\rm max}\approx 6.5\times 10^4~{\rm K}~[({\dot M}/{\dot M}_{\rm Edd})/(M/10^8~M_\odot)]^{0.07}$ dependence found analytically by \citet{2014MNRAS.438.3024L}.  We see that wind losses weaken the dependence of maximum disk temperature as a function of outer accretion rate, as is consistent with several recent observations of optically detected TDE candidates \citep{2012Natur.485..217G,2014ApJ...780...44C,2014MNRAS.445.3263H}. 
 
\section{RESULTS}
\label{sec:results}

\begin{figure}[!htb]
\begin{center}
\plotone{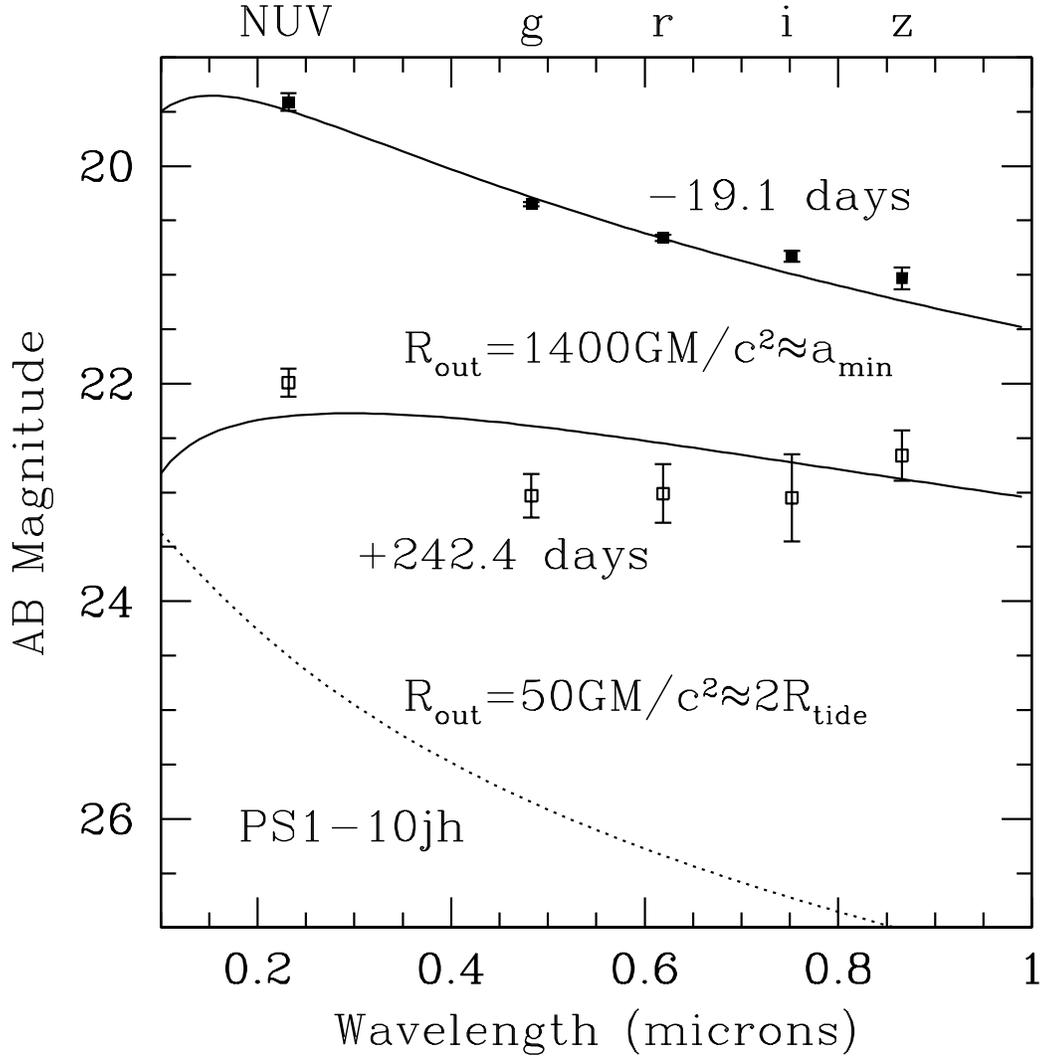}
\vskip-1.5cm
\caption{Illustrative fit of the disk wind model to the magnitude data for the candidate TDE PS1--10jh given in \citet{2012Natur.485..217G}; the solid symbols with error bars are for 19.1 days before maximum luminosity, and the open symbols with error bars are for 242.4 days after maximum luminosity.  See text for details of the model.  The solid lines are for an outer disk radius that is comparable to the semimajor axis of the most bound debris, whereas the dotted line shows the output of the model assuming an outer disk radius that is twice the tidal radius.  Other models fit equally well (e.g., a standard disk model with no wind, or even a blackbody as shown by \citealt{2012Natur.485..217G}), but this figure demonstrates that both the fluxes and the colors require that the emitting area must be closer to the area characteristic of the most bound debris than to the area at the nominal circularization radius (see also \citealt{2014arXiv1410.7772S}).
}
\label{fig:fit10jh}
\end{center}
\end{figure}

In this section we display illustrative fits of our model to data from two optically detected TDEs.  In Figure~\ref{fig:fit10jh}, we show an example comparison of the ${\dot\Sigma}(F)$, $\epsilon=0$ model to data from the PS1--10jh TDE taken from \citet{2012Natur.485..217G}.  Here we assume a nonrotating black hole with a mass of $M=3\times 10^6~M_\odot$ (other masses give comparably good fits), and we have incorporated the redshift $z=0.1696$ and the implied luminosity distance $d_{\rm lum}=820$~Mpc to this source.  The dotted line shows the disk wind model spectrum for a disk with an outer radius comparable to the circularization radius for a stream with a pericenter equal to the tidal radius.  For this line we assume an accretion rate at the outer edge that is 15 times the Eddington accretion rate for a black hole of this mass that accretes pure hydrogen.  Clearly the optical and ultraviolet fluxes are far too small, and the colors are far too blue, compared with the data.  In contrast, the solid lines show disk wind models with an outer radius $R_{\rm out}=1400~GM/c^2$, which is comparable to the semimajor axis of the most bound debris.  For the data 19.1 days before maximum, we assume an outer disk accretion rate that is 15 times the Eddington accretion rate for pure hydrogen, and for the data 242.4 days after maximum, we assume an outer accretion rate that is 0.25 times the Eddington rate for pure hydrogen. These models are obviously far superior to the small-disk models, although the agreement with the data is not perfect.  

\begin{figure}[!htb]
\begin{center}
\plotone{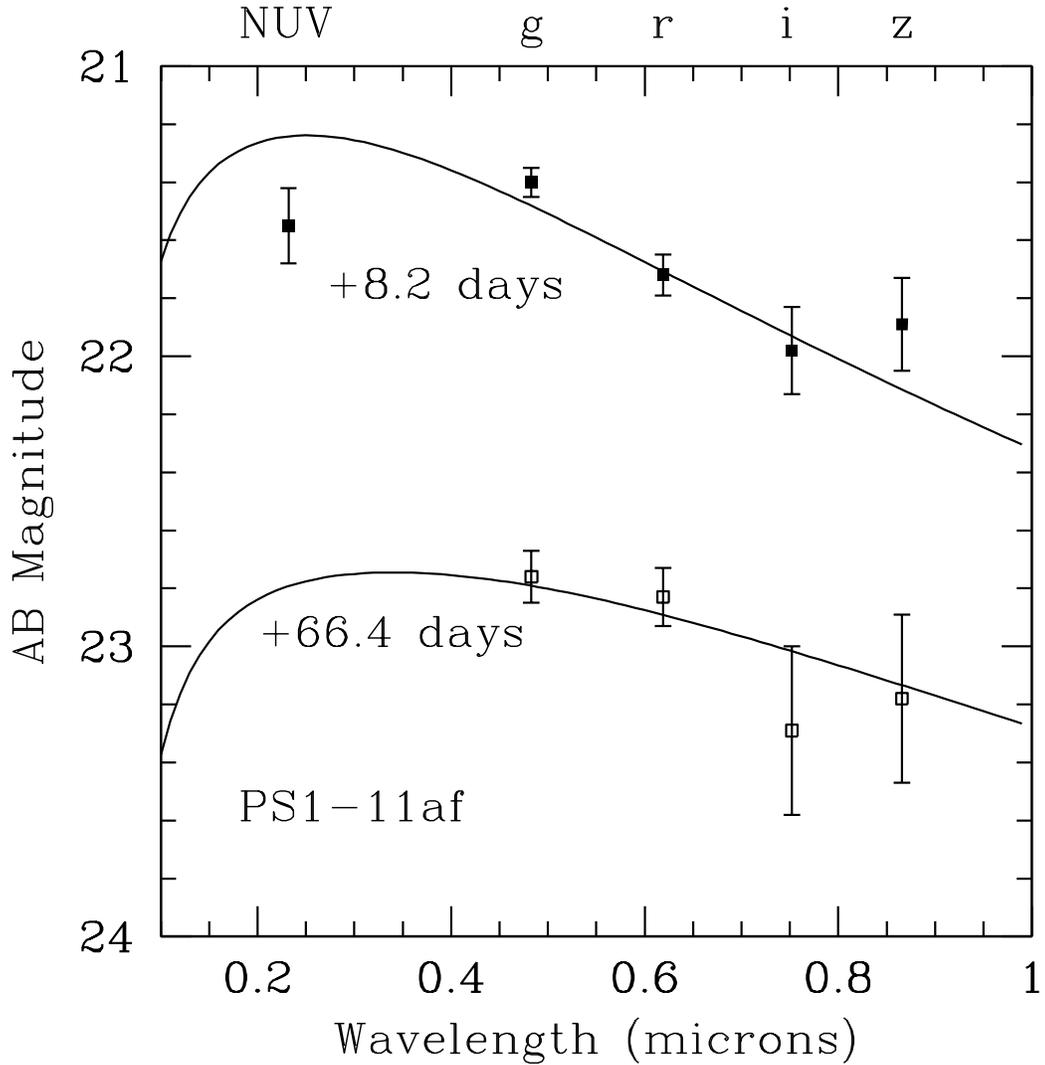}
\vskip-1.5cm
\caption{Same as Figure~\ref{fig:fit10jh}, but for the candidate TDE PS1--11af \citep{2014ApJ...780...44C}.  The data are again labeled by time relative to maximum luminosity.  For both lines the outer disk radius is $3000GM/c^2$.  The outer accretion rate for the data 8.2 days after maximum is 14 times the Eddington rate for a pure hydrogen disk around a $M=3\times 10^6~M_\odot$ black hole, and for the data 66.4 days after maximum is 1.8 times the Eddington rate.  Again, the disk wind model provides reasonable fits to the spectral data.
}
\label{fig:fit11af}
\end{center}
\end{figure}

Similarly, we show in Figure~\ref{fig:fit11af} a by-eye fit of the ${\dot\Sigma}(F)$, $\epsilon=0$ model to data from the PS1--11af TDE taken from \citet{2014ApJ...780...44C}.  We again assume a nonrotating black hole with $M=3\times 10^6~M_\odot$ (and again, other masses would work as well), and we have incorporated the redshift $z=0.4046$ and the implied luminosity distance $d_{\rm lum}=2200$~Mpc to this source.  For the data at 8.2 days after maximum, we assume an outer disk accretion rate at the outer edge that is 14 times the Eddington accretion rate for pure hydrogen.  For the data from 66.4 days after maximum, we assume an outer disk accretion rate that is 1.8 times the Eddington accretion rate.  In both cases, our outer disk radius is $R_{\rm out}=3000GM/c^2$.  The fits to PS1--10jh and PS1--11af suggest that the heat from stream interactions cools inefficiently \citep{2015arXiv150104635B,2015arXiv150105306G,2015arXiv150105207H,2015arXiv150104365S}.

The time scale of the evolution of the fallback rate implied by our fit to the PS1--10jh data is approximately consistent with models.  For example, Figure~21 of \citet{2015arXiv150104365S} shows the ballistic fallback rate scaled from their simulations (this corresponds to the outer disk accretion rate in our model).  The unit of time is the orbital time at the outer disk radius, which for our example fit is $2\pi(R_{\rm out}^3/GM)^{1/2}=56$~days for $R_{\rm out}=1400~GM/c^2$ and $M=3\times 10^6~M_\odot$.  The implied rise time of $\sim 50$~days is comparable to what \citet{2012Natur.485..217G} found for PS1--10jh.  The scale $M_*/(GM/c^3)$ for the accretion rate used by \citet{2015arXiv150104365S} is $3.13\times 10^{31}~{\rm erg~s}^{-1}$ if $M_*=0.23~M_\odot$ \citep{2012Natur.485..217G}, so at 19 days before the peak, or $\sim 0.3$ time units before the peak, the expected accretion rate would be $\approx 2\times 10^{-6}~M_*/(GM/c^3)\approx 6.26\times 10^{25}~{\rm g~s}^{-1}$, which is roughly 10 times the hydrogen Eddington accretion rate onto a $3\times 10^6~M_\odot$ nonrotating black hole.  At 242 days after the peak, or approximately 4.3 time units after the peak, the expected fallback rate would be roughly 10 times less than it was 19 days before the peak rather than the 60 times less that we find in our model.  

\begin{figure}[!htb]
\begin{center}
\plotone{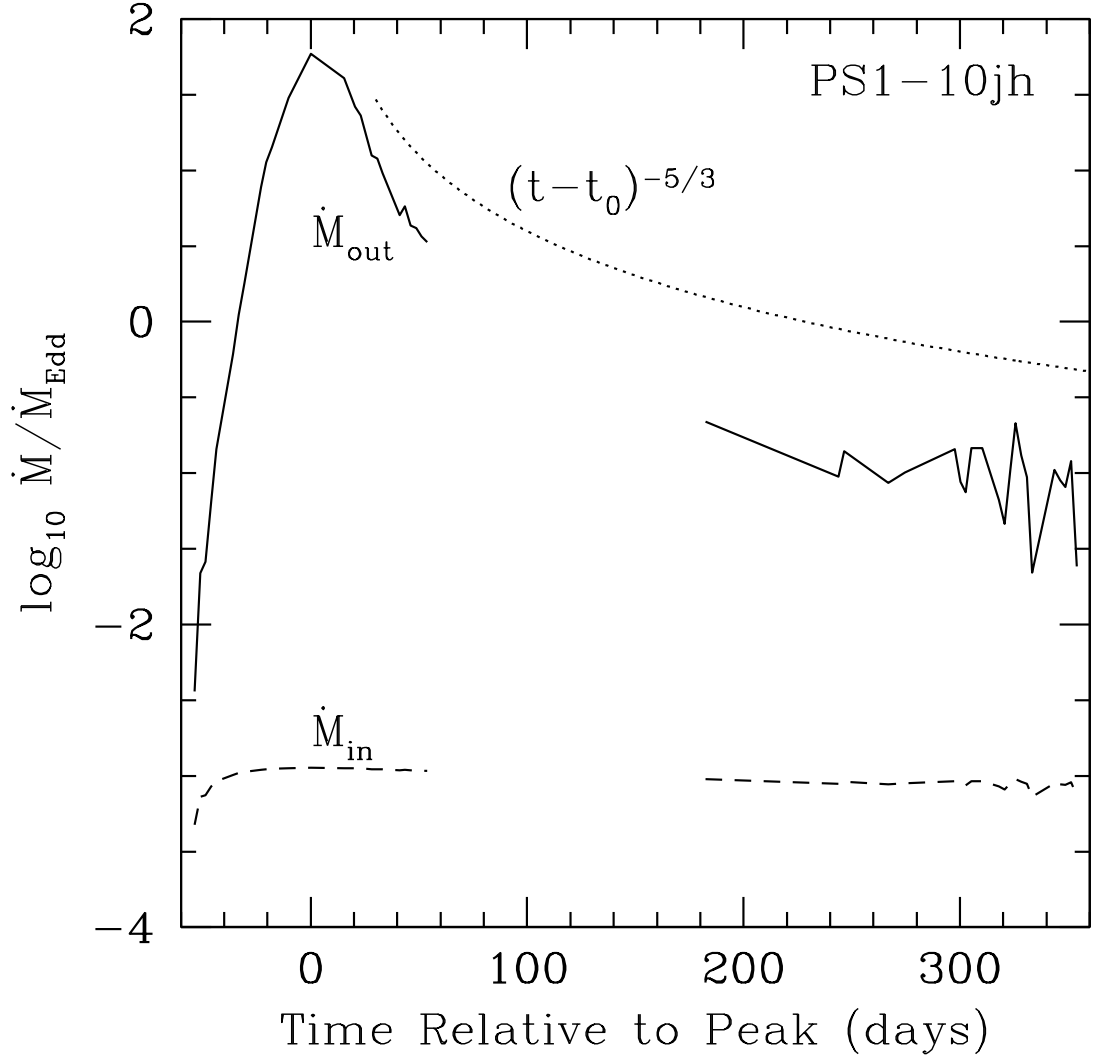}
\vskip-1.5cm
\caption{Mass accretion rate at the outer edge of the disk (${\dot M}_{\rm out}$, labeled with the solid line) and into the black hole (${\dot M}_{\rm in}$, labeled with the dashed line), for a particular model of the PS1--10jh event in which ${\dot\Sigma}={\dot\Sigma}(F)$, $\epsilon=0$, $M=3\times 10^6~M_\odot$, and $R_{\rm out}=1400~GM/c^2$ for all fits.  We obtained ${\dot M}$ by finding the accretion rate that gave a g-band magnitude equal to the best estimate in \citet{2012Natur.485..217G}; beyond $\sim 240$~days the large photometric uncertainties lead to choppiness in this best estimate.  There were no data from $\sim50$ to $\sim180$ days after the peak.  For comparison we also plot an offset curve that shows the relative accretion rate for a $(t-t_0)^{-5/3}$ law.  Caveats apply regarding the simplicity of the model, but the inferred fallback rate is comparable to what is expected.
}
\label{fig:mdotfit}
\end{center}
\end{figure}

In Figure~\ref{fig:mdotfit} we show an example of the outer and inner accretion rate inferred using a particular model of the PS1--10jh event.  Here we assume ${\dot\Sigma}={\dot\Sigma}(F)$, $\epsilon=0$, $M=3\times 10^6~M_\odot$, and $R_{\rm out}=1400~GM/c^2$ for all data, and fit only to the best-estimate g-band magnitude without extinction corrections given in \citet{2012Natur.485..217G}.  There are many reasons why the disk structure could differ from our assumptions (for example, as was recently pointed out by \citealt{2015arXiv150205792P}, the disk could remain elliptical for many orbits), but the general agreement with a ${\dot M}_{\rm out}\sim (t-t_0)^{-5/3}$ law is encouraging.  In this model the integrated mass that fell on the outer edge of the disk over the $\sim 1$~yr of observation was $M_{\rm out}\approx 1.5~M_\odot$, but the integrated mass that fell into the black hole was only $\approx 2\times 10^{-3}~M_\odot$.  Similar masses are inferred for ${\dot\Sigma}={\dot\Sigma}(F,g)$ or $\epsilon=1$, and lower masses ($M_{\rm out}$ equal to a few tenths of a solar mass) are inferred for larger outer radii.  The implied fallback mass is plausible for a stellar disruption, but the overwhelming majority of the matter is driven away in a wind.  This wind will carry a considerable amount of kinetic energy $E_{\rm kin}\sim M_\odot v^2\sim 10^{50}$~erg; whether, and how, this energy becomes manifest will likely depend on the strength of interactions the wind has with any surrounding gas in the galactic center.

The time scale implied by our fit to the PS1--11af event does not fit as well as it does to the PS1--10jh data.  From the data of \citet{2014ApJ...780...44C} the rise time of the event is shorter by tens of percent than it is for PS1--10jh, but our model requires a larger disk and therefore a larger characteristic time.  Perhaps in this source the self-intersection of the tidal stream proceeds in a slightly different way.  For example, it could be that the shocks due to stream-self-intersection in PS1--11af are particularly strong; this drives the initial inflow, as was demonstrated by \citet{2015arXiv150104365S}.  Thus for PS1--10jh our model provides a reasonable fit to the spectra and to the time scales, whereas for PS1--11af our model provides a reasonable fit to the spectra but is off by a moderate factor from the observed time scales.  

Many caveats apply to such fits.  For example, there is an unknown level of extinction intrinsic to the host galaxy, there is an unknown amount of contamination from other sources, and the smoothness of the ultraviolet and optical data means that many models give equally good formal fits (this includes disk models without mass loss to winds and, as is shown in Figure~3 of \citealt{2012Natur.485..217G}, even simple Planck spectra).  There are also likely to be other sources of variation in mass accretion rate with radius, especially if the inflow time is long compared with the mass accumulation time at the outer edge of the disk.  In addition, for this fit we assume for simplicity that each disk annulus emits as a blackbody, whereas in reality there will be spectral corrections introduced by radiative transfer (e.g., \citealt{2005ApJ...621..372D}).  What we highlight here is what we believe to be a robust result: in order to explain the ultraviolet and optical data it is necessary that the characteristic disk size be considerably greater than than the $R_{\rm out}=2r_p$ that comes from straightforward angular momentum conservation (see also \citealt{2014arXiv1410.7772S}).  This is true even relatively early in the disruption, which suggests that the disk becomes large quickly.

\section{CONCLUSIONS}
\label{sec:conclusions}

We have shown that two aspects of optically discovered TDEs can be explained qualitatively by a model involving accretion disk winds that was originally proposed for AGN by \citet{2014MNRAS.438.3024L}.  These aspects are the low maximum temperature from TDE disks and the weak evolution of that temperature as a function of the fallback rate.  In both cases the idea is that wind losses throughout the disk act to regulate the accretion rate through the inner portion of the disk.  Therefore, increasing the accretion rate at the outer edge of the disk increases the inner disk temperature by a far smaller factor than would be expected in the standard no-wind disk model.  This implies that massive black holes are messy eaters of stars, in the sense that the accretion rate is much lower than the supply rate from the disruption; for example, as we showed in Section~\ref{sec:results}, it could be that well under 1\% of the infalling matter is actually accreted by the black hole.  Thus consumption of stars may be a highly inefficient way of growing massive black holes.  In addition, we find suggestive evidence in two TDEs that the characteristic disk size rapidly exceeds $2r_p$, which would be the expected circularization radius, and approaches the size $a_{\rm min}$ of the most bound orbit of the debris.  This is consistent with recent simulations in which cooling of the shocked debris is inefficient \citep{2015arXiv150104635B,2015arXiv150105306G,2015arXiv150105207H,2015arXiv150104365S}, and in particular consistent with the specific suggestion of \citet{2015arXiv150104365S} that $R_{\rm out}\sim a_{\rm min}$.  One might, in this model, expect that the wind would have an effect on the profiles of broad emission lines (e.g., \citealt{2012ApJ...753..133F} show that such lines become narrower with increasing wind optical depth), but a detailed investigation of the implications for lines is beyond the scope of this paper.

A natural question is: under what circumstances can the maximum disk temperature be high enough that it produces detectable X-rays, as is required for X-ray discovered TDEs?  There are several possibilities, none of which is likely to suffice in every case.  First, in some circumstances, it could be that the matter is not sufficiently thermalized to emit as a blackbody, but instead radiates less efficiently and thus reaches a higher temperature (as could happen in coronae or jets, or in the thermal bremsstrahlung suggestion of \citealt{1993MNRAS.262..141S}). Second, given that line-driven winds require heavy elements, if the metallicity is low then less mass is driven away and greater temperatures can be achieved. Third, as in stellar-mass black hole binaries, there could be a threshold temperature beyond which the ionization fraction is great enough that the relevant opacity is Thomson rather than the opacity given by overlapping metal lines; this would become progressively more important at lower black hole masses, so X-ray detection could sometimes indicate lower mass primaries.  If this is the explanation then there could be circumstances in which, as the fallback rate drops and the temperature decreases, line opacities become more important and there is a rapid drop in X-ray flux as the disk temperature decreases.  Finally, on occasion,  variability of the accretion onto black holes could be mistaken for a TDE.

In summary, disk winds offer a natural explanation for the low temperatures and weak dependence of temperature on mass accretion rate that is seen in several optically-discovered TDEs.  If this explanation is correct, it suggests that observations of TDEs can probe non-standard disk models as well as the self-interaction of streams after the initial disruption.

\acknowledgements

This work was supported in part by NASA ATP grant NNX12AG29G.  This work was also supported in part by the National Science Foundation under Grant No. PHYS-1066293 and the hospitality of the Aspen Center for Physics.  We thank Tamara Bogdanovi\'c, Roseanne Cheng, Jane Dai, Suvi Gezari, James Guillochon, Julian Krolik, Ari Laor, Enrico Ramirez-Ruiz, Elena Rossi, Hotaka Shiokawa, and Nick Stone for helpful suggestions prior to the submission of this manuscript.  We also thank the anonymous referee for an unusually constructive report.

\bibliography{wind}

\end{document}